\documentclass{article}
\usepackage{amssymb,amsfonts,amsbsy,amsthm}

\newtheorem{theorem}{Theorem}
\newtheorem{corollary}{Corollary}
\newtheorem{proposition}{Proposition}
\theoremstyle{definition}
\newtheorem{definition}{Definition}
\newtheorem{example}{Example}

\newcommand{\K}{\mathbb{K}}
\newcommand{\F}{\mathbb{F}}
\newcommand{\Q}{\mathbb{Q}}
\newcommand{\C}{\mathbb{C}}

\begin{document}

\title{Computing the Fixing Group of a Rational Function}

\author{Jaime Gutierrez \and Rosario Rubio \and David Sevilla}

\date{}
\maketitle

\begin{abstract}
Let $\mathrm{Aut}_\K\K(x)$ be the Galois group of the
transcendental degree one pure field extension $\K\subseteq\K(x)$.
In this paper we describe polynomial time algorithms for computing
the field $Fix(H)$ fixed by a subgroup
$H\subseteq\mathrm{Aut}_\K\K(x)$ and for computing the fixing
group $G_f$ of a rational function $f\in\K(x)$.
\end{abstract}

\section{Introduction}

Let $\K$ be an arbitrary field and $\K(x)$ be the rational
function field in the variable $x$. Let $\mathrm{Aut}_\K\K(x)$ be
the Galois group of the field extension $\K\subseteq\K(x)$.

In this paper we develop an algorithm for computing the
automorphism group of an intermediate field in the extension
$\K\subseteq\K(x)$. By the classical L\"uroth's theorem any
intermediate field $\F$ between $\K$ and $\K(x)$ is of the form
$\F=\K(f)$ for some rational function $f\in\K(x)$, see
\cite{Lur,Sch} and for a constructive proof \cite{GRS01}. Thus,
this computational problem is equivalent to determine the fixing
group $G_f$ of a univariate rational function $f$. We also present
an algorithm for computing $Fix(H)$, the fixed field by a subgroup
$H\subseteq\mathrm{Aut}_\K\K(x)$. Again, this computational
problem is equivalent to finding a L\"uroth's generator of the
field fixed by the given subgroup $H$. Both algorithms are in
polynomial time if the field $\K$ has a polynomial time algorithm
for computing the set of the roots of a univariate polynomial.

The algorithm for computing the fixing group of a rational
function uses several techniques related to the rational function
decomposition problem. This problem can be stated as follows:
given $f\in\K(x)$, determine whether there exists a decomposition
$(g,h)$ of $f$, $f=g(h)$, with $g$ and $h$ of degree greater than
one, and in the affirmative case, compute one. When such a
decomposition exists some problems become simpler: for instance,
the evaluation of a rational function $f$ can be done with fewer
arithmetic operations, the equation $f(x)=0$ can be solved more
efficiently, improperly parametrized algebraic curves can be
reparametrized properly, etc. see \cite{Zip}, \cite{nearsep} and
\cite{Sed}. In fact, a motivation of this paper is to obtain
results on rational functional decomposition. As a consequence of
our study of $G_f$ we provide new and interesting conditions of
decomposability of rational functions. Another application of this
paper is to study the number $m$ of indecomposable components of a
rational function $f=f_1\circ\cdots\circ f_m$ which is strongly
related to the group $G_f$, see \cite{GRS02}.

The other algorithm presented for computing the field $Fix(H)$ is based on Galois
theory results and the constructive proof of L\"uroth's theorem.

The paper is divided in four sections. In  Section 2 we introduce
our notations and the background of rational function
decomposition. Section 3 studies the Galois group of $\K(x)$ over
$\K$, the fixing group $G_f$ and the  field $Fix(H)$, including
general theoretical results, and its relation with the functional
decomposition problem. Section 4 presents algorithms for computing
the fixing group and fixed field. We also give in this section
examples illustrating our algorithms.

\section{Background on rational function decomposition}

The set of all non--constant rational functions is a semigroup
with identity $x$, under the element--wise composition of rational
functions (symbol $\circ$ for composition): i.e., given
non--constant rational functions $g,h\in\K(x)$, $g\circ h=g(h)$.
The units of this semigroup are of the form $\frac{ax+b}{cx+d}$.
We will identify these units with the elements of the Galois group
of $\K(x)$ over $\K$. We will denote this group by
$\Gamma(\K)=\mathrm{Aut}_\K\K(x)$.

Given $f\in\K(x)$, we will denote as $f_N,f_D$ the numerator and
denominator of $f$ respectively, assuming that $f_N$ and $f_D$ are
relatively prime. We define the degree of $f$ as $\deg
f=\max\{\deg f_N,\deg f_D\}$.

If $g,h\in\K(x)$ are rational functions of degree greater than
one, $f=g\circ h=g(h)$ is their (functional) composition, $(g,h)$
is a (functional) decomposition of $f$, and $f$ is a decomposable
rational function.

The following lemma describes some basic properties of rational
function decomposition, see \cite{nearsep} for a proof.

\begin{theorem}
With the above notations and definitions, we have the following:
\begin{itemize}
\item $[\K(x):\K(f)]=\deg f$.
\item $\deg g\circ h=\deg g\cdot\deg h$.
\item The units with respect to composition are precisely the
rational functions $u$ with $\deg u=1$.
\item Given $f,h\in\K(x)\setminus\K$, if there exists $g$ such
that $f=g(h)$, it is unique. Furthermore, it can be computed from
$f$ and $h$ by solving a linear system of equations.
\end{itemize}
\end{theorem}

If $f,h\in\K(x)$ satisfy $\K(f)\subset\K(h)\subset\K(x)$, then
$f=g(h)$ for some $g\in\K(x)$. From this fact the following
natural concept arises:

\begin{definition}
Let $f=g\circ h=g'\circ h'$. $(g,h)$ and $(g',h')$ are called \textbf{equivalent
decompositions} if there is a unit $u$ such that $h'=u\circ h$ (then also
$g'=g\circ u^{-1}$).
\end{definition}

The next result is an immediate consequence of L\"uroth's theorem.
\begin{corollary}
Let $f\in\K(x)$ be a non--constant rational function. Then the equivalence
classes of the decompositions of $f$ correspond bijectively to intermediate
fields $\F$, $\K(f)\subseteq\F\subset\K(x)$.
\end{corollary}

\section{The Galois Correspondences in the Extension $\K\subseteq\K(x)$.}

We start defining our main notions and tools.

\begin{definition} Let $\K$ be any field.
\begin{itemize}
\item Let $f\in\K(x)$. The \textbf{fixing group} of $f$ is $G_f$,
$$G_f=\{u\in\Gamma(\K): f\circ u=f\}.$$
\item Let $H$ be a subgroup of $\Gamma(\K)$. The \textbf{fixed field}
by $H$ is $\mathrm{Fix}(H)$,
$$\mathrm{Fix}(H)=\{f\in\K(x): f\circ u=f\ \forall u\in H\}.$$
\end{itemize}
\end{definition}

Before we discuss the computational aspects of these concepts, we
will need some properties based on general facts from Galois
theory.

\begin{theorem}\label{propiedad}$ $
\begin{itemize}
\item Let $H<\Gamma(\K)$.
\begin{itemize}
  \item $H$ is infinite $\Rightarrow \mathrm{Fix}(H)=\K$.
  \item $H$ is finite $\Rightarrow \K\varsubsetneq \mathrm{Fix}(H)$,
  $\mathrm{Fix}(H)\subset\K(x)$ is a normal extension, and in
  particular $\mathrm{Fix}(H)=\K(f)$ with $\deg f=|H|$.
\end{itemize}

\item Given a finite subgroup $H$ of $\Gamma$, there is a
bijection between the subgroups of $H$ and intermediate fields in
$\mathrm{Fix}(H)\subset\K(x)$. Also, if $\mathrm{Fix}(H)=\K(f)$,
there is a bijection between the right components of $f$ (up to
equivalence by units) and the subgroups of $H$.

\item Given $f\in\K(x)\setminus\K$, the
order of $G_f$ divides $\deg f$. Moreover, for every $\K$ there is an $f\in\K(x)$
such that $1<|G_f|<\deg f$, for example if $f=x^2\,(x-1)^2$ then $G_f$=\{x,1-x\}.

\item If $|G_f|=\deg f$ then the extension $\K(f)\subset\K(x)$ is
normal. Moreover, if the extension $\K(f)\subset\K(x)$ is also separable, then
$\K(f)\subset\K(x)$ is normal implies $|G_f|=\deg f$.

\item $G_f$ depends on the field $\K$: let
$f=x^4$, then for $\K=\Q, G_f=\{x,-x\}$ but for $\K=\Q(i), G_f=\{x,-x,ix,-ix\}$.

\item If $\K$ is infinite, then $f\in\K \Leftrightarrow G_f$ is infinite.

\item Let $f\in\K(x)$ and $u,v$ be two units. Let $H<\Gamma(\K)$
\begin{itemize}
\item Let  If $f'=u\circ f\circ v$, then
$G_f=v\cdot G_{f'}\cdot v^{-1}$.
\item If $\mathrm{Fix}(H)=\K(f)$ then for any $\alpha$, $\mathrm{Fix}(\alpha H\alpha^{-1})=\K(f\circ\alpha^{-1})$.
\end{itemize}

\item It is possible that $f$ is decomposable but
$G_f$ is trivial: for $\K=\C, f=x^2(x-1)^2(x-3)^2$; for $\K=\Q, f=x^9$.

\item It is possible that $f$ has a non-trivial decomposition $f=g(h)$
and $G_f$ is not trivial but $G_h$ is not a proper subgroup of $G_f$: for $\K=\C,
f=(x^2-1)(x^2-3) \Rightarrow G_f=\{x,-x\}$; for $\K=\Q, f=x^4 \Rightarrow
G_f=\{x,-x\}$.
\end{itemize}
\end{theorem}

Unfortunately, it is not true that $[\K(x):\K(f)]=|G_f|$. However, some
interesting results about decomposability can be given.

\begin{theorem}\label{cond}
Let $f$ be indecomposable.
\begin{itemize}
\item If $\deg f=p$ is prime, then either $G_f\cong C_p$ or
$G_f$ is trivial.
\item If $\deg f=n$ is not prime, then $G_f$ is trivial.
\end{itemize}
\end{theorem}

According of above theorem if $H$ is infinite then $Fix(H)$ is trivial. Some
times it is interesting to see the element of $\Gamma(\K)$ as matrices.

\begin{proposition}\label{isom}
The group $\Gamma(\K)$ is isomorphic to $PGL_2(\K)=GL_2(\K)/D_2(\K)$ where
$D_2(\K)=\{\lambda I_2:\lambda\in\K^*\}$. Moreover, if $\K$ is algebraically
closed, then it is also isomorphic to $PSL_2(\K)=SL_2(\K)/\{\pm I_2\}$.
\end{proposition}

The study of the finite subgroups of $\Gamma(\C)$ has a long
history. Any finite subgroup corresponds to the a rotation or
reflection of the Riemann sphere, so the finite subgroups
correspond to the regular solids in three dimensions. Klein
\cite{klein} gave the first geometric proof of the following
classification of the finite subgroups of $\Gamma(\C)$.

\begin{theorem}\label{klein}[Klein]
Every finite subgroup of $\Gamma(\C)$ is isomorphic to one of the
following groups:
\begin{itemize}
\item $C_n$, the cyclic group of order $n$;
\item $D_n$, the dihedral group of order $n$;
\item $A_4$, the alternating group on four letters or \textsl{tetrahedral} group;
\item $S_4$, the symmetric group on four letters or \textsl{octahedral} group;
\item $A_5$, the alternating group on five letters or \textsl{icosahedral} group.
\end{itemize}
\end{theorem}

In the case $\K=\Q$, the correspondence between functions and
groups is not so good as in the complex case, see Theorem
\ref{propiedad}. On the other hand, it is not difficult (personal
communication of Prof. Walter Feit) to obtain from Theorem
\ref{klein} a classification of all finite subgroups of
$\Gamma(\Q)$.

Now, suppose that $\K$ is finite, that is, $\K=\F_q$ where $q$ is
a power of a prime $p$, $q=p^n$. We denote the set of all linear
polynomials with coefficients in $\F_q$ by
$\Gamma_0(\F_q)=\{ax+b:a\in\F_q^*,b\in\F_q\}$.

\begin{theorem}\label{finito} With the above notation, we have the following:
\begin{itemize}
\item $|\Gamma_0(\F_q)|=q^2-q$, $|\Gamma(\F_q)|=q^2-q$.
\item $\Gamma_0(\F_q)<\Gamma(\F_q)$, but $\Gamma_0(\F_q)\ntriangleleft\Gamma(\F_q)$.
\item The group $\Gamma(\F_q)$ is generated by $\Gamma_0(\F_q)$ and the linear
rational function $1/x$, that is, $\Gamma=<\Gamma_0,1/x>$.
\item $Fix(\Gamma_0(\F_q))=\F_q(f_0)$, where $f_0=(x^q-x)^{q-1}$.
\item $Fix(\Gamma(\F_q))=\F_q(h(f_0))$, where $h=\frac{x^{q+1}+x+1}{x^q}$
\end{itemize}
\end{theorem}

As a consequence of Theorem \ref{finito} we have the following
theoretical result:

\begin{theorem}
For every $\K$, the extension $\K\subset\K(x)$ is Galois if and
only if $\K$ is infinite.
\end{theorem}

\section{Algorithms}

Now, we have all ingredients to give a computational solution to
both problems.

\subsection{Algorithm for computing the fixed field}

As the next theorem shows, it is easy to compute a generator for
the fixed field of an explicitly given group (suggested by Dr.
Peter M\"uller).

\begin{theorem}\label{Muller}
Let $G=\{g_1,\ldots,g_m\}\subseteq\K(x)$ be a finite group. Let
$P(t)=\prod_1^m (t-g_i)\in\K(x)[t]$. Then any non--constant
coefficient of $P(t)$ generates $F_G$.
\end{theorem}

The following example illustrates the algorithm over the field
$\C$.

\begin{example}
Let \[G=\{ \pm\frac{t-i}{t+i},\pm\frac{t+i}{t-i},\pm\frac{1}{t},\pm
t,\pm\frac{i(t-1)}{t+1},\pm\frac{i(t+1)}{t-1}\}<\Gamma(\C)\] which is isomorphic
to $A_4$. All the symmetric functions in the elements of $G$ are in
$\mathrm{Fix}(G)$, and any non--constant symmetric function generates it. We
compute those functions:
\begin{itemize}
\item $\sigma_1=\sigma_3=\sigma_5=\sigma_7=\sigma_9=\sigma_{11}=0$ by
symmetry in the group.
\item
$\sigma_2=\sigma_{10}=\displaystyle{\frac{-1+33t^4+33t^8-t^{12}}{t^{10}-2t^6
+t^2}}$.
\item $\sigma_4=\sigma_8=\displaystyle{\frac{-33t^4-66t^2-33}{t^4+2t^2+1}}$.
\item
$\sigma_6=\displaystyle{\frac{2-66t^4-66t^8+2t^{12}}{t^2-2t^6+t^{10}}}$.
\item $\sigma_{12}=1$.
\end{itemize}
\end{example} \qed

\subsection{Algorithm for computing the fixing group}

The most straightforward method of computing the fixing group of a rational
functions is solving a polynomial system of equations. Given
\[f=\frac{a_nx^n+\cdots+a_0}{b_mx^m+\cdots+b_0}\]
we have the system given by equating to 0 the coefficients of the numerator of
$f\circ\left(\frac{ax+b}{cx+d}\right)-f(x)$. We can alternatively solve the two
systems given by \newline $f\circ(ax+b)-f(x)=0$ and
$f\circ\left(\frac{ax+b}{x+d}\right)-f(x)=0$. This method is simple but
inefficient; we will present another method that is faster and will allow us to
extract useful information even if the group is not computed completely.

We will assume that $\K$ has sufficiently many elements; if it is not the case,
we can work in an extension and check later which elements are in $\Gamma(\K)$.

\begin{definition}
Let $f\in\K(x)$. We say $f$ is in \textbf{normal form} if $\deg f_N>\deg f_D$ and
$f_N(0)=0$.
\end{definition}

\begin{theorem}
Let $f\in\K(x)$. If $\K$ has sufficiently many elements, there exist units $u$
and $v$ such that $u\circ f\circ v$ is in normal form.
\end{theorem}

\begin{theorem}
Let $f\in\K(x)$ be in normal form and $u=\frac{ax+b}{cx+d}$ such that $f\circ
u=f$.
\begin{description}
\item $a\neq 0$ and $d\neq 0$.
\item $f_N(b/d)=0$.
\item If $c=0$, that is $u=ax+b$, then $f_N(b)=0$ and $a^n=1$ where $n=\deg
f$.
\item If $c\neq 0$ then $f_D(a/c)=0$.
\end{description}
\end{theorem}

In order to compute $G_f$, we use the previous theorem to compute
the polynomial and rational units separately.

Thus, if we can compute the roots of any polynomial in $\K[x]$, we
have the following algorithm:

\noindent \textbf{Algorithm}
\begin{description}
\item[Input:] $f\in\K(x)$.
\item[Output:] $G_f=\{u\in\K(x): f\circ u=f\}$.
\item [A.] Compute units $u,v$ such that $f'=u\circ f\circ v$ is in
normal form. Let $n=\deg f$. Let $L$ be an empty list.
\item [B.] Compute $A=\{\alpha\in\K: \alpha^n=1\}$, $B=\{\beta\in\K: f'_N(\beta)=0\}$ and $C=\{\gamma\in\K: f'_N(\gamma)=0\}$..
\item [C.] For each $(\alpha,\beta)\in A\times B$, check if
$f'\circ(ax+b)=f'$. If that is the case, add $ax+b$ to $L$.
\item [D.] For each $(\beta,\gamma)\in B\times C$, let $w=\frac{c\gamma x+\beta}{cx+1}$
and compute all values of $c$ for which $f'\circ w=f'$. For each solution, add
the resulting unit to $L$.
\item [E.] Let $L=\{w_1,\ldots,w_k\}$. Then, RETURN $\{v\circ w_i\circ
v^{-1}: i=1,\ldots,k\}$.
\end{description}

\begin{example}
Let \[f={\frac {\left(-3\,x+1+x^3\right)^2}
{x\left(-2\,x-x^2+1+x^3\right)\left(-1+x\right)}} \in\Q(x).\]
First we normalize $f$: let $u=1/(x-9/2)$ and $v=1/x-1$, then
\[f'=u\circ f\circ
v=\frac{-4x^6-6x^5+32x^4-34x^3+14x^2-2x}{27x^5-108x^4+141x^3-81x^2+21x-2}\]
is in normal form.

The roots of the numerator and denominator of $f'$ in $\Q$ are $\{0,1,1/2\}$ and
$\{1/3,2/3\}$ respectively. The only sixth roots of unity in $\Q$ are $1$ and
$-1$; as $\mathrm{char\ }\Q=0$ there are no elements of the form $x+b$ in
$G_{f'}$. Therefore, there are two polynomial candidates to test: $-x+1/3$ and
$-x+2/3$. It is easy to check that none of them leaves $f$ fixed.

Let $w=\frac{c\beta x+\alpha}{cx+1}$.
\begin{description}
\item [$\alpha=0,\beta=1/3:$] the unit $\frac{cx/3}{cx+1}$ does
not leave $f$ fixed for any value of $c$.
\item [$\alpha=1,\beta=1/3:$] the unit $\frac{cx/3+1}{cx+1}$ does
not leave $f$ fixed for any value of $c$.
\item [$\alpha=1/2,\beta=1/3:$] the unit $\frac{cx/3+1/2}{cx+1}$ leaves $f$ fixed for $c=-3/2$.
\item [$\alpha=0,\beta=2/3:$] the unit $\frac{2cx/3}{cx+1}$ does
not leave $f$ fixed for any value of $c$.
\item [$\alpha=1,\beta=2/3:$] the unit $\frac{2cx/3+1}{cx+1}$ leaves $f$ fixed for $c=-3$.
\item [$\alpha=1/2,\beta=2/3:$] the unit $\frac{2cx/3+1/2}{cx+1}$ does
not leave $f$ fixed for any value of $c$.
\end{description}
Therefore,
\[G_{f'}=\{x,\frac{-x+1}{-3x+2},\frac{-2x+1}{-3x+1}\}\]
and
\[G_f=v\cdot G_{f'}\cdot v^{-1}=\{x,\frac{1}{1-x},\frac{x-1}{x}\}.\]

From this group we can compute a proper component of $f$ using
Theorem \ref{Muller}, an we obtain
\[h=\frac{-3\,x+1+x^3}{\left(-1+x\right)x}\]
which is indeed a right--component for $f$, since $f=g\circ h$ with
\[g=\frac{x^2}{x-1}.\]
\end{example}  \qed

Now we present an example illustrating the algorithm over a finite
field.

\begin{example}
Let $\K=\F_2$ and \[f=\frac{(x^2+1)(x^6+x^4+x^2+1+x^3)}
{x^8+x^4+1+x^5+x^3}.\]

First we normalize $f$: let $u=\frac{x+1}{x}$ and $v=\frac{1}{x}+1$. Then
\[f'=u\circ f\circ v=\frac{(x+1)^4 x^4}{(x^2+x+1)(x^4+x+1)}.\]

As $B=\{0,1\}$ and $C=\emptyset$, we only have to check the unit
$x+1$. As it leaves $f'$ fixed, we have that $G_{f'}=\{x,x+1\}$
and
\[G_f=v\cdot G_{f'}\cdot v^{-1}=\{x,\frac{1}{x}\}.\]
Therefore, a generator of $\mathrm{Fix}(G_f)$ is
\[h=x+\frac{1}{x}\]
which is also a component of $f$: $f=g\circ h$ with
\[g=\frac{x^4+x}{x^4+x+1}.\]
\end{example} \qed

\noindent \textbf{Acknowledgments}

\noindent This research is partially supported by the National
Spanish project PB97-0346 and Grant Movilidad del Personal de
Investigador.

\end{document}